\documentclass{llncs}
\pdfoutput=1
\usepackage[pdftex]{graphicx}
\usepackage{float}
\usepackage{url}
\usepackage{color}
\usepackage{makecell}

\def\omp{OpenMP}
\def\acc{OpenACC}
\def\cuda{CUDA}
\def\ssoln{\textbf{save\_soln}}
\def\adt{\textbf{adt\_calc}}
\def\res{\textbf{res\_calc}}
\def\bres{\textbf{bres\_calc}}
\def\update{\textbf{update}}
\usepackage{array}
\newcolumntype{P}[1]{>{\centering\arraybackslash}p{#1}}
\newcolumntype{M}[1]{>{\centering\arraybackslash}m{#1}}

\begin{document}

\title{Comparison of Parallelisation Approaches, Languages, and Compilers for Unstructured Mesh Algorithms on GPUs}
\author{G. D. Balogh\inst{1} \and I. Z. Reguly\inst{1}
\and G. R. Mudalige\inst{2}}

\authorrunning{Balogh et al.} 
%
\tocauthor{G. D. Balogh, I. Z. Reguly, G. R. Mudalige}
\institute{Faculty of Information Technology and Bionics, Pazmany Peter Catholic University, Budapest, Hungary,\\
\email{balogh.gabor.daniel@hallgato.ppke.hu, reguly.istvan@itk.ppke.hu},
\and
Department of Computer Science, University of Warwick, Coventry, \\ 
United Kingdom \email{g.mudalige@warwick.ac.uk}}

\maketitle              

\begin{abstract}

Efficiently exploiting GPUs is increasingly essential in scientific computing, as many current and upcoming supercomputers are built using them. To facilitate this, there are a number of programming approaches, such as CUDA, OpenACC and OpenMP 4, supporting different programming languages (mainly C/C++ and Fortran). There are also several compiler suites (clang, nvcc, PGI, XL) each supporting different combinations of languages. In this study, we take a detailed look at some of the currently available options, and carry out a comprehensive analysis and comparison using computational loops and applications from the domain of unstructured mesh computations. Beyond runtimes and performance metrics (GB/s), we explore factors that influence performance such as register counts, occupancy, usage of different memory types, instruction counts, and algorithmic differences. Results of this work show how clang's CUDA compiler frequently outperform NVIDIA's nvcc, performance issues with directive-based approaches on complex kernels, and OpenMP 4 support maturing in clang and XL; currently around 10\% slower than CUDA.

\end{abstract}

\section{Introduction}

The last ten years has seen the widespread adoption of Graphical Processing Units (GPUs) by the high performance computing community. For a wide range of highly parallel workloads they offer higher performance and efficiency. Programming techniques for GPUs have also evolved significantly. The CUDA \cite{cuda} language extensions to C/C++ and the OpenCL language \cite{opencl} provide a low-level programming abstraction commonly referred to as Single Instruction Multiple Thread (SIMT) that gives fine-grained control over GPU architectures. CUDA/OpenCL allows the exploitation of low-level features like scratch pad memory, warp operations, and block-level synchronization. However, converting existing applications to use CUDA or OpenCL is a substantial undertaking that require significant effort and considrable changes to the design of the programe and the source code. Furthermore, getting good performance can entail detailed work in orchestrating parallelism.

To simplify the adoption of GPUs, particularly for existing codes, high-level directive based programming abstractions were introduced. OpenACC \cite{openacc} introduced in 2011 was one of the first supporting GPUs. Subsequetly OpenMP standard introduced support for accelerators starting from version 4 \cite{openmp4}, with refinements in 4.5 and 5.0. Of particular note is that the evolution of directive based approaches being driven by the acquisition of large US DoE systems such as Titan and the upcoming Summit and Sierra systems. To be able to efficiently utilize these systems it was necessary that existing codes be modified to support GPUs with relative ease. Many of these codes are written in Fortran and as such there is now compiler support for writing CUDA, OpenACC, and OpenMP with Fortran in various compilers.

It is generally agreed that the best performance can be achieved by using CUDA, but the difference between CUDA and directive-based approaches vary significantly based on a multitude of factors. Primarily these include the type of computation being parallelized, as well as the language being used (C or Fortran), and the compiler. This motivates the present study: for a number of parallel loops, coming from the domain of unstructured mesh computations, we wanted to get an idea of what performance looks like on different GPUs, different languages, and different compilers. Given the available systems and compilers, we would like to acertain what the state-of-th-art is with regard to utilizing GPU based systems for this class of applications. 

We evaluate some of the most commonly used compilers and parallelization approaches. We explore the performance of CUDA C, compiled with nvcc, as well as with Google's recent clang based compiler \cite{Wu:2016:GOG:2854038.2854041}. We also explore the performance of the compilers by Portland Group (PGI, now owned by NVIDIA) which has had support for wirting CUDA applications in Fortran \cite{pgi,ruetsch2013cuda}. Additionally, as part of a recent push by IBM, preparing for the Summit and Sierra machines there has been support for CUDA Fortran with the XL compilers since v15.1.5 \cite{xlcuf}. We also explore XL compiler performance in this paper. For OpenACC we use the PGI compilers which support both C and Fortran. There is also good support for OpenACC by the Cray compilers, however we did not have access to such a machine and therefore will not be part of this analysis. For OpenMP 4 there are two compilers developed by IBM directed at developing applications using C: the XL compilers (since v13.1.5), and an extension to Clang \cite{clang-ykt}. There is also support for writing OpenMP 4 parallizations in Fortran applications using the XL compilers (since v15.1.5).

While there is a tremendous amount of research on performance evaluation of various combinations of languages and compilers, we believe our work is unique in its breadth: it directly compares C and Fortran implementations of the same code (Airfoil), and with three different parallelizations: CUDA, OpenACC, and OpenMP, and with five different state-of-the-art compilers. We also present an in-depth study trying to explain the differences with the help of instruction counters and the inspection of low-level code. Specifically, we make the following contributions:
\begin{enumerate}
\item Using a representative CFD application called Airfoil, we run the same algorithms on NVIDIA K40 and P100 GPUs, with CUDA, OpenMP 4, and OpenACC parallelizations written in both C and Fortran, compiled with a number of different compilers.
\item We carry out a detailed analysis of the results with the help of performance counters to help identify differences between algorithms, languages, and compilers.
\item We evaluate these parallelizations and compilers on two additional applications, Volna (C) and BookLeaf (Fortran) to confirm the key trends and differences observed on Airfoil.
\end{enumerate}

The rest of the paper is structured as follows: Section \ref{sec/related} discusses some related work, Section \ref{sec/apps} briefly introduces the applications being studied, then Section \ref{sec/test} presents the test setup, compilers and flags. Section \ref{sec/bench} carries out the benchmarking of parallelizations and the detailed analysis, and finally Section \ref{sec/conc} draws conclusions.

\section{Related Work} \label{sec/related}

There is a significant body of existing research on performance engineering for GPUs, and compiler engineering, as well as some comparisons between parallelization approaches - the latter however is usually limited in scope due to the lack of availability of multiple implementations of the same code. Here we cite some examples, to show how this work offers a wider look at the possible combinations.

Work by Ledur et. al. compares a few simple testcases such as  Mandelbrot and N-Queens implemented with CUDA and OpenACC (PGI) \cite{ledur2013comparative}, Herdman et. al. \cite{herdman2012accelerating} take a larger stencil code written in C, and study CUDA, OpenCL and OpenACC implementations, but offer no detailed insights into the differences. Work by Hoshino et. al. \cite{6546071} offers a detailed look at CUDA and OpenACC variants of a CFD code and some smaller benchmarks written in C, and show a few language-specific optimizations, but analysis stops at the measured runtime. Normat et. al. \cite{norman2015case} compare CUDA Fortran and OpenACC versions of an atmospheric model, CAM-SE, which offers some details about code generated by the PGI and Cray compilers, and identifies a number of key differences that let CUDA outperform OpenACC, thanks to lower level optimizations, such as the use of shared memory. Kuan et. al.  \cite{kuan2014accelerating} also compare runtimes of CUDA and OpenACC implementations of the same statistical algorithm (phylogenetic inference). Gonge et. al. \cite{gong2016nekbone} compare CUDA Fortran and OpenACC implementations of Nekbone, and scale up to 16k GPUs on Titan - but no detailed study of performance differences.

Support in compilers for OpenMP 4 and GPU offloading is relatively new \cite{antao2016offloading} and there are only a handful of papers evaluating their performance: Martineau et. al. \cite{Martineau2016} present some runtimes of basic computational loops in C compiled with Cray and clang, and  comparisons with CUDA. Karlin et. al \cite{Karlin2016} port three CORAL benchmark codes to OpenMP 4.5 (C), compile them with clang, and compare them with CUDA implementations - the analysis is focused on runtimes and register pressure. Hart el. al. \cite{Hart2015FirstEP} compare OpenMP 4.5 with Cray to OpenACC on Nekbone, however the analysis here is also restricted to runtimes, the focus is more on programmability. We are not aware of academic papers studying the performance of CUDA Fortran or OpenMP 4 in the IBM XL compilers aside from early results in our own previous work \cite{reguly2016high}. There is also very little work on comparing the performance of CUDA code compiled with nvcc and clang.

Thus we believe that there is a significant gap in current research: a comparison of C and Fortran based CUDA, OpenACC, and OpenMP 4, the evaluation of the IBM XL compilers, the maturity of OpenMP 4 compared to CUDA in terms of performance and a more detailed investigation into the reasons for the performance difference between various languages, compilers, and parallelization approaches. With the present study, we work towards filling this gap.

\vspace{-10pt}
\section{Applications} \label{sec/apps}
\vspace{-10pt}

The applications being studied in this work come from the unstructured mesh computations domain solving problems in the areas of computational fluid dynamics, shallow-water 
simulation and Lagrangian hydrodynamics. As such, they consist of parallel loops over some set in the mesh, such as edges, cells or nodes, and on each set element some computations are carried out, while accessing data either directly on the iteration set, or indirectly via a mapping to another set. Our applications are all written using the OP2 domain specific language \cite{op2} embedded in C and Fortran, targeting unstructured mesh computations. For OP2, the user has to give a high level description of the simulation using the OP2 API. Then the OP2 source-to-source translator generates all parallelized versions from the abstract description \cite{mudalige2012op2}. While OP2 is capable of many things, its relevant feature for this work is that it can generate different parallelizations such as CUDA, OpenACC, and OpenMP4, based on the abstract description of parallel loops.

\begin{figure}
\includegraphics[width=0.9\linewidth]{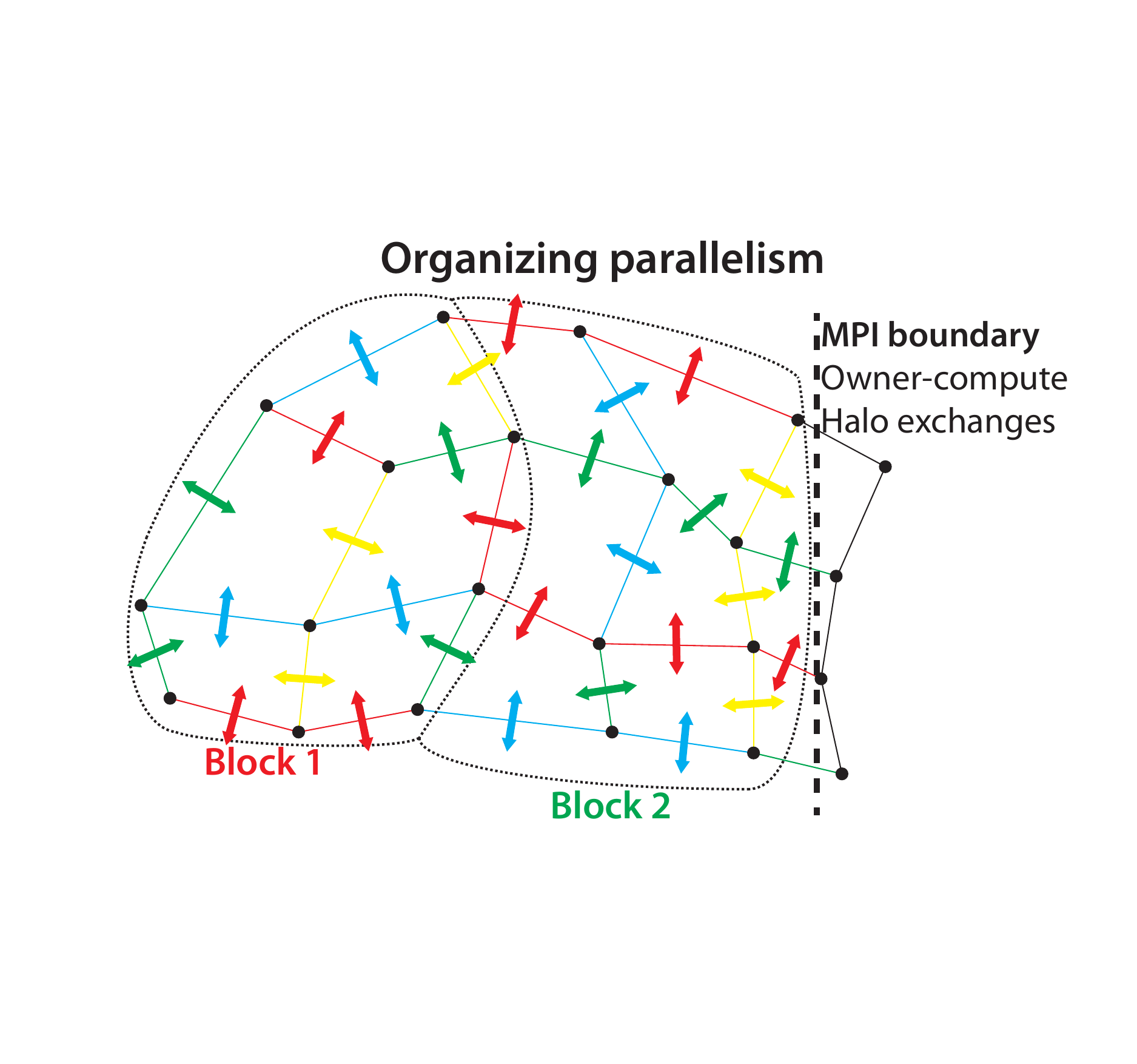}
\caption{Illustration for hierarchical coloring on a computation on edges that write data on the cells. The blocks are colored so that there is no neighboring blocks with the same color and inside the blocks threads colored so that no two threads with the same color write the same data.}
\label{fig:coloring}
\end{figure}

A key challenge in unstructured mesh computations is the handling of race conditions when data is indirectly written. For the loops with indirect increments (which means we incrementing some value through a mapping so there are multiple iterations incrementing the same value), we use coloring to ensure that no two threads will write to the same memory at the same time. We can use a more sophisticated coloring approach for GPUs using CUDA as described in \cite{giles2011performance}, where we create and color mini-partitions such that no two mini-partitions of the same color will update the same cell. This allows mini-partitions of the same color to be processed by the blocks of one \cuda\ kernel. Within these mini-partitions, each assigned to a different CUDA thread block, each thread will process a different element within these blocks, and thus is it necessary to introduce a further level of coloring. For an edges to cells mapping, we color all edges in a mini-partition so that no two edges with the same color update the same cell. Such a coloring is shown in figure \ref{fig:coloring}. Here, we first calculate the increment of every thread in the block, then we iterate through the colors and add the increment to the cell with synchronization between each color. The benefit of such an execution scheme is that there is a possibility that the data we loaded from the global memory can be reused within a block, which can lead to a performance increase due to fewer memory transactions. This technique is referred to as hierarchical coloring in the paper.

With other methods such as OpenACC and OpenMP4 there is no method for thread synchronization and data sharing in blocks, which is essential for the hierarchical coloring technique described above. Therefore a global coloring technique is used in case of these parallelization approaches. This technique is similar to the thread coloring inside the mini-partitions, but works on the full set. We assign colors to each thread in a way that no two edges of the same color update the same cell and threads from the same color can run parallel in a separate CUDA kernel with synchronization between the kernels. This however excludes the possibility of the reuse of the data of the cells.

\subsection{Airfoil}\vspace{-5pt}
Airfoil is a benchmark application, representative of large industrial CFD applications. It is a non-linear 2D inviscid airfoil code that uses an unstructured grid and a finite-volume discretisation to solve the 2D Euler equations using a scalar numerical dissipation. The algorithm iterates towards the steady state solution, in each iteration using a control volume approach, meaning the change in the mass of a cell is equal to the net flux along the four edges of the cell, which requires indirect connections between cells and edges. Airfoil is implemented using OP2, where two versions exists, one implemented with OP2's C/C++ API and the other using OP2's Fortran API \cite{giles2012op2,op2}.

The application consists of five parallel loops: \textbf{save\_soln}, \textbf{adt\_calc}, \textbf{res\_calc}, \textbf{bres\_calc} and \textbf{update} \cite{mudalige2012op2}. The \textbf{save\_soln} loop iterates through cells and is a simple loop accessing two arrays directly. It basically copies every four state variables of cells from the first array to the second one.
The \textbf{adt\_calc} kernel also iterates on cells and it computes the local area/timestep for every single cell. For the computation it reads values from nodes indirectly and writes in a direct way. There are some computationally expensive operations (such as square roots) performed in this kernel.
The \textbf{res\_calc} loop is the most complex loop with both indirect reads and writes; it iterates through edges, and computes the flux through them. It is called 2000 times during the total execution of the application and performs about 100 floating-point operations per mesh edge.
The \textbf{bres\_calc} loop is similar to \textbf{res\_calc} but computes the flux for boundary edges.
Finally \textbf{update} is a direct kernel that includes a global reduction which computes a root mean square error over the cells and updates the state variables.
\\
\indent All test are executed with double precision on a mesh containing 2.8 million cells and with SOA data layout described in \cite{mudalige2012op2}.

\vspace{-10pt}\subsection{Volna}\vspace{-5pt}
Volna is a shallow water simulation capable of handling the complete life-cycle of a tsunami (generation, propagation and run-up along the coast) \cite{dutykh2011volna}. The simulation algorithm works on unstructured triangular meshes and uses the finite volume method. Volna is written in C/C++ and converted to use the OP2 library\cite{op2}. For Volna we examined the top three kernels where most time is pent: \textbf{computeFluxes}, \textbf{SpaceDiscretization} and \textbf{NumericalFluxes}. In the \textbf{computeFluxes} kernel there are indirect reads and direct writes, in \textbf{NumericalFluxes} there are indirect reads with direct writes and a global reduction for calculating the minimum timestep and in \textbf{SpaceDiscretization} there are indirect reads and indirect increments. \\ 
\indent Tests are executed in single precision, on a mesh containing 2.4 million triangular cells, simulating a tsunami run-up to the US pacific coast.

\vspace{-10pt}\subsection{BookLeaf}\vspace{-5pt}
BookLeaf is a 2D unstructured mesh Lagrangian hydrodynamics application from the UK Mini-App Consortium \cite{uk-mac}. It uses a low order finite element method with an arbitrary Lagrangian-Eulerian method. Bookleaf is written entirely in Fortran 90 and has been ported to use the OP2 API and library. Bookleaf has a large number of kernels with different access patterns such as indirect increments similar to increments inside \res\ in Airfoil. For testing we used the SOD testcase with a 4 million element mesh. We examined the top five kernels with the highest runtimes which are \textbf{getq\_christiensen1}, \textbf{getq\_christiensen\_q}, \textbf{getacc\_scatter}, \textbf{gather}, \textbf{getforce\_visc}. Among these there is only one kernel (\textbf{getacc\_scatter}) with indirect increments (where coloring is needed), the \textbf{gather} and \textbf{getq\_christiensen1} have indirect reads and direct writes as \adt\ in Airfoil, and the other two kernels have only direct reads and writes.

\vspace{-10pt}\section{Test setup} \label{sec/test}\vspace{-10pt}
For testing we used NVIDIA K40 and P100 GPUs in IBM S824L  systems (both systems has 2*10 cores) with Ubuntu 16.04. We used nvcc in CUDA 9.0 and clang 6.0.0 (r315446) for compiling \cuda\ with C/C++. For compiling CUDA Fortran, we used PGI 17.4 compilers and IBM's XL compiler 15.1.6 beta 12 for Power systems. For \omp4, we tested clang version 4.0.0 (commit 6dec6f4 from the clang-ykt repo), and the XL compilers (13.1.6 beta 12). Finally, for OpenACC, we used the PGI compiler version 17.4. The specific compiler versions and flags are shown in Table \ref{tab:compiler_flags}. \\

\vspace{-25pt}
\begin{table}
	\begin{tabular}{M{.15\linewidth}|M{.09\linewidth}|M{.76\linewidth}}
&Version&Flags\\ \hline
PGI				 &17.4-0&-O3 -ta=nvidia,cc35 -Mcuda=fastmath -Minline=reshape (-acc for OpenACC)\\ \hline
XL				 &15.1.6 beta 12 13.1.6 beta 12&-O3 -qarch=pwr8 -qtune=pwr8 -qhot -qxflag=nrcptpo -qinline=level=10 -Wx,-nvvm-compile-options=-ftz=1 -Wx,-nvvm-compile-options=-prec-div=0 -Wx,-nvvm-compile-options=-prec-sqrt=0 (-qsmp=omp -qthreaded -qoffload for \omp4)\\ \hline
clang for OpenMP4&4.0	&-O3 -ffast-math -fopenmp=libomp -Rpass-analysis -fopenmp-targets=nvptx64-nvidia-cuda -fopenmp-nonaliased-maps -ffp-contract=fast\\ \hline
clang for CUDA	 &6.0	&-O3 --cuda-gpu-arch=sm\_35 -ffast-math\\ \hline
nvcc			 &9.0.176&-O3 -gencode arch=compute\_35,code=sm\_35 --use\_fast\_math \\ \hline
	\end{tabular}
    \caption{Compiler flags used on K40 GPU (for P100 cc60 and sm\_60 is used)}
    \label{tab:compiler_flags}
\end{table}
\vspace{-45pt}

\vspace{-5pt}\section{Benchmarking} \label{sec/bench}\vspace{-10pt}

\subsection{Airfoil}\vspace{-10pt}

\begin{figure}
	\centering
		   	\vspace{-20pt}
	   	\includegraphics[width=0.80\textwidth]{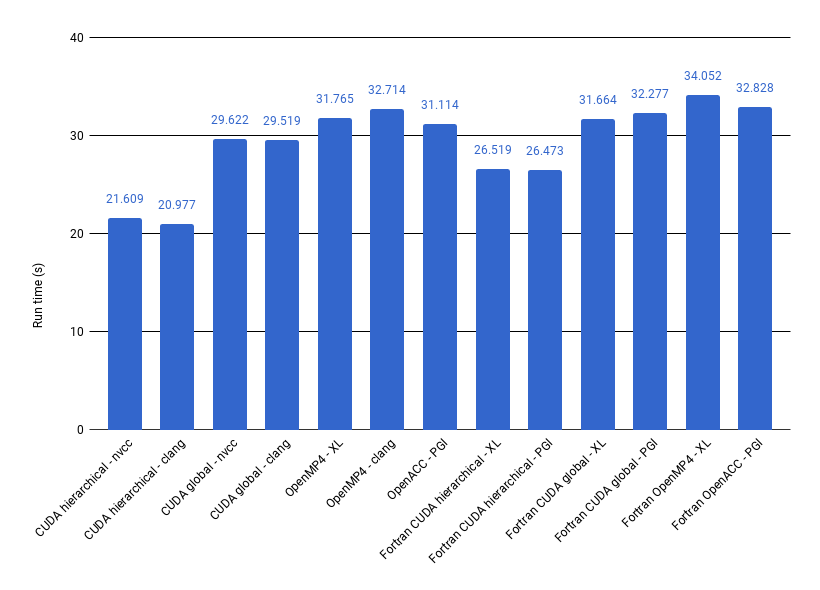}
        \vspace{-20pt}
    	\includegraphics[width=0.80\textwidth]{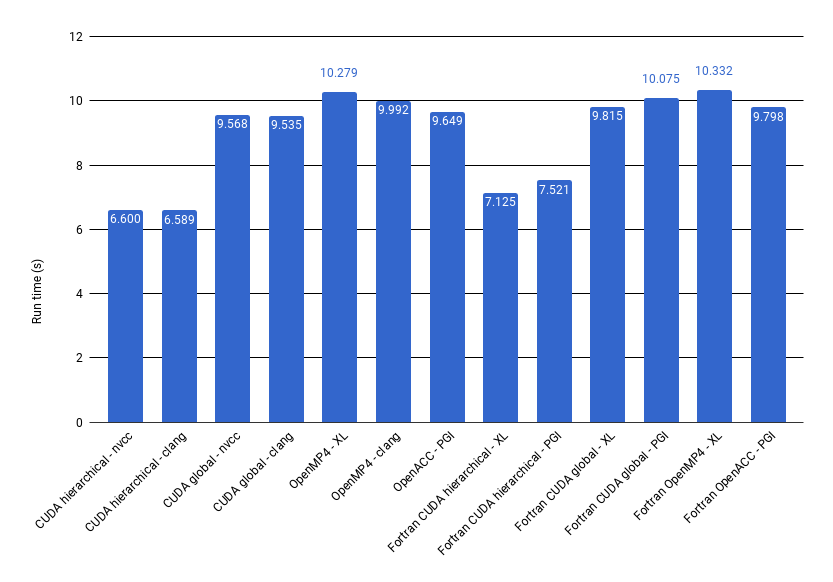}
	    \caption{Measured run times of versions on the K40 and P100 GPU\vspace{-30pt}}
	    \label{fig:P100_airfoil_RT}
\end{figure}

The run times of different versions of Airfoil on the K40 and P100 GPUs are shown in Figure \ref{fig:P100_airfoil_RT}. The hierarchical coloring is used in \res\ and \bres, because these have indirect increments and in the case of other kernels we don't need coloring because they have only direct updates. The versions using the hierarchical coloring scheme have the best performance, due to the huge performance gains in \res\ thanks to data reuse. The main differences between versions with the same coloring strategy is in the run times of the \res\ and \adt\ kernels, where most of the computation is performed. In the following, we examine performance in detail on all five kernels.
\renewcommand{\floatpagefraction}{.8}%
\vspace{-20pt}

\subsubsection{\ssoln}:
\ssoln\ is a really simple kernel with only direct reads and writes. It copies state variables of the cells, and thus is highly memory bounded. In \cuda\ versions we used 200 blocks so each thread processes more than one cell to save on integer instructions. However this leads us to a for loop inside the kernel, increasing control instructions, and slowing performance. In Table \ref{tab:save_soln} the runtimes of the \ssoln\ kernel are shown: all versions have approximately the same performance. The bandwidth values shown in the table are the useful bandwidth from the users perspective, that is the sum of the moved simulation data and mappings for the kernel divided by the run time of the kernel. In case of C/C++, \omp4 and \acc\ versions have about $5-7\%$ better runtimes and bandwidth than \cuda\ versions (even though the \omp4 version compiled with clang on the K40 GPU is the only version that have only 75\% occupancy). If we run one thread per cell and delete the loop from the kernel the performance of \cuda\ matches the performance of \omp4. The results shows that in a simple case such as \ssoln\ Fortran performs about as well as the C/C++ versions, and the high level approaches such as \omp4 and \acc\ can reach the performance of \cuda.
\vspace{-15pt}
\begin{table}
    \centering
    \begin{tabular}{l|P{1.25cm}|P{1.05cm}|P{1.9cm}|P{1.25cm}|P{1.05cm}|P{1.9cm}}
    						& \multicolumn{3}{c}{K40} 		& \multicolumn{3}{c}{P100}\\
         					& Run Time (s) & BW (GB/s) & Reg. count (Occupancy) & Run Time (s) & BW (GB/s) & Reg. count (Occupancy)\\ \hline
         nvcc - \cuda		&1.055&175&21 (100\%)& 0.362&509&24 (100\%)\\ \hline
         clang - \cuda		&1.055&175&21 (100\%)& 0.362&509&24 (100\%)\\ \hline
         PGI - \acc			&1.006&183&26 (100\%)& 0.351&526&29 (100\%)\\ \hline
         XL - \omp4			&1.003&184&17 (100\%)& 0.357&517&19 (100\%)\\ \hline
         clang - \omp4		&0.982&188&35 (75\%) & 0.356&518&32 (100\%)\\ \hline\hline
         PGI - F\_\cuda		&1.061&174&32 (100\%)& 0.368&502&32 (100\%)\\ \hline
         PGI - F\_\acc		&1.012&182&24 (100\%)& 0.349&528&29 (100\%)\\ \hline
         XL - F\_\cuda		&1.060&174&32 (100\%)& 0.362&502&32 (100\%)\\ \hline
         XL - F\_\omp4		&1.009&183&22 (100\%)& 0.352&524&24 (100\%)\\ \hline
    \end{tabular}
    \caption{Measured run time, bandwidth, register count and occupancy values in case of \ssoln}\vspace{-20pt}
    \label{tab:save_soln}
\end{table}

\subsubsection{\adt}:

In case of \adt\ the loop iterates over cells and reads data indirectly from the nodes while updating a single value per cell multiple times. The operation contains some expensive square root calculations which introduce high numbers of additional floating point operations and increase the register counts for the kernel. 
For \adt\ the directive based approaches use significantly higher numbers of registers than \cuda~as shown in Table~\ref{tab:adt_calc}; this means lower occupancy and about 30\% worse performance on the K40 machine (on the P100 machine the difference is only about 10-20\%). In case of \omp4 with the XL compiler and \acc\ with the PGI compiler every time the value on the cell is written we see a global store instruction instead of calculating the intermediate results in registers and only write the final results to the global memory as other versions do. Another source of performance difference for \omp4 with clang compiler comes from the lack of usage of texture caches for loading the read only data from the nodes. The cause of clang \cuda\ slightly outperforming nvcc \cuda\ on the K40 machine is that it computes the expensive square root operations with fewer floating point instructions, which leads to about 16\% less floating point instructions than nvcc (this also holds for the P100 card) for \adt.
The Fortran versions have high register counts, thus lower occupancy, this is one of the key reasons for the 30\% lower performance on the K40 GPU (on the P100 GPU the difference is about 10-15\%), also Fortran versions use about 50\% more integer instructions than C/C++ versions. The directive based approaches perform within 20\% of \cuda\ Fortran's performance and the with the PGI compiler the versions execute twice as many integer instructions than other versions.

\begin{table}
    \centering
    \begin{tabular}{l|P{1.25cm}|P{1.05cm}|P{1.9cm}|P{1.25cm}|P{1.05cm}|P{1.9cm}}
    						& \multicolumn{3}{c}{K40} 		& \multicolumn{3}{c}{P100}\\
         					& Run Time (s) & BW (GB/s) & Reg. count (Occupancy) & Run Time (s) & BW (GB/s) & Reg. count (Occupancy)\\ \hline
         nvcc - \cuda		& 2.810&148&40 (75\%)   & 0.869 &477&40 (75\%)   \\ \hline
         clang - \cuda		& 2.756&151&36 (75\%)   & 0.867 &478&40 (75\%)   \\ \hline
         PGI - \acc			& 4.071&102&86 (31.25\%)& 0.978 &424&96 (31.25\%)\\ \hline
         XL - \omp4			& 3.775&110&64 (50\%)   & 0.984 &421&72 (43.75\%)\\ \hline
         clang - \omp4		& 4.108&101&88 (31.25\%)& 1.077 &385&96 (31.25\%)\\ \hline\hline
         PGI - F\_\cuda		& 3.753&116&64 (50\%)   & 0.955 &434&56 (56.25\%)\\ \hline
         PGI - F\_\acc		& 4.341&96 &86 (31.25\%)& 1.053 &394&96 (31.25\%)\\ \hline
         XL - F\_\cuda		& 3.581&116&78 (37.5\%) & 1.001 &415&88 (31.25\%)\\ \hline
         XL - F\_\omp4		& 3.905&106&80 (37.5\%) & 1.090 &380&86 (31.25\%)\\ \hline
    \end{tabular}
    \caption{Measured run time, bandwidth, register count and occupancy values in case of \adt}\vspace{-30pt}
    \label{tab:adt_calc}
\end{table} 

\subsubsection{\res}:
In the case of \res\ we have indirect updates, therefore we need coloring to avoid race conditions. The runtime and bandwidth results are shown in Table \ref{tab:res_calc_2l} for hierarchical coloring, and for global coloring in Table \ref{tab:res_calc}.
In this kernel there is a lot of indirectly read and written data, therefore the runtime can be significantly improved with the hierarchical coloring approach due to data reuse. However, hierarchical coloring leads to higher register counts and arithmetic instruction counts, but the impact of these factors are smaller than the gain from better memory usage. As we saw it in \adt\ for \cuda\ versions clang performs better than nvcc in terms of integer and floating point instruction counts (clang has 2-5\% lower instruction counts on both GPUs). The \omp4 and \acc\ versions have 2-5\% higher run time because of low occupancy (caused by register pressure) and the \omp4 versions don't use the texture caches as much (or at all in case of clang) as other versions which lead to 3 times as much global loads and high number of integer instructions. The results are shown in Table \ref{tab:res_calc_2l_instruction} and Table \ref{tab:res_calc_glob_instruction}.\\
Fortran versions with hierarchical coloring have 23-30\% worse performance than the same C/C++ versions, while with global coloring this difference is only 5\% (15\% on P100), but generally Fortran versions have high register counts and lower occupancy, as well as higher numbers of integer instructions and global load transactions. With Fortran the differences between the performance of \cuda\ and directive based approaches are about the same as described above.\vspace{20pt}

\begin{table}
    \centering
    \begin{tabular}{l|P{1.25cm}|P{1.05cm}|P{1.9cm}|P{1.25cm}|P{1.05cm}|P{1.9cm}}
    						& \multicolumn{3}{c}{K40} 		& \multicolumn{3}{c}{P100}\\
         					& Run Time (s) & BW (GB/s) & Reg. count (Occupancy) & Run Time (s) & BW (GB/s) & Reg. count (Occupancy)\\ \hline
         nvcc - \cuda	& 13.118&67&53 (56.25\%)& 3.727&235&50 (56.25\%)\\ \hline
         clang - \cuda	& 12.537&70&56 (56.25\%)& 3.721&235&51 (56.25\%)\\ \hline\hline
         PGI - F\_\cuda	& 16.880&72&69 (43.75\%)& 4.425&198&78 (37.5\%)\\ \hline
         XL - F\_\cuda	& 16.235&54&72 (43.75\%)& 3.968&221&70 (43.75\%)\\ \hline
    \end{tabular}
    \caption{Measured run time, bandwidth, register count and occupancy values of \res\ in case of hierarchical coloring}
    \label{tab:res_calc_2l}\begin{tabular}{l|P{1.25cm}|P{1.05cm}|P{1.9cm}|P{1.25cm}|P{1.05cm}|P{1.9cm}}
    						& \multicolumn{3}{c}{K40} 		& \multicolumn{3}{c}{P100}\\
         					& Run Time (s) & BW (GB/s) & Reg. count (Occupancy) & Run Time (s) & BW (GB/s) & Reg. count (Occupancy)\\ \hline
         nvcc - \cuda		& 21.133&41&46 (62.5\%)& 6.706&131&40 (56.25\%)\\ \hline
         clang - \cuda		& 21.083&42&46 (62.5\%)&6.676&131&40 (56.25\%)\\ \hline
         PGI - \acc			& 21.472&41&72 (43.75\%)&6.617&132&88 (31.25\%)\\ \hline
         XL - \omp4			& 22.277&39&71 (43.75\%)& 7.200&122&80 (37.5\%)\\ \hline
         clang - \omp4		& 22.245&39&96 (31.25\%)& 6.676&131&96 (31.25\%)\\ \hline\hline
         PGI - F\_\cuda		& 22.700&38&87 (31.25\%)& 6.993&125&88 (31.25\%)\\ \hline
         PGI - F\_\acc		& 22.992&38&87 (31.25\%)& 6.713&130&96 (31.25\%)\\ \hline
         XL - F\_\cuda		& 22.236&39&88 (31.25\%)& 6.806&129&94 (31.25\%)\\ \hline
         XL - F\_\omp4		& 23.755&37&110 (25\%)  & 7.229&121&104 (25\%)\\ \hline
    \end{tabular}
    \caption{Measured run time, bandwidth, register count and occupancy values of \res\ in case of global coloring}
    \label{tab:res_calc}
\end{table}

\begin{table}
    \centering
    \begin{tabular}{l|c|c|c|c}
         									 & nvcc   & clang& Fortran PGI & Fortran XL \\ \hline
        integer instructions 				 & 191743K& 0.86 &  0.82& 0.90 \\ \hline
        floating point (64 bit) instructions & 88698K & 0.87 &  0.97& 0.95 \\ \hline
        Control instructions				 & 8955K  & 0.90 &  0.64& 0.26 \\ \hline
        Texture read transactions 			 & 761K   & 1.00 & 16.91& 8.88 \\ \hline
        Global read Transactions 			 & 188K   & 1.00 &  0.95& 4.22 \\ \hline
    \end{tabular}
    \caption{Average number of instructions and transactions performed in \res\ kernel with hierarchical coloring (absolute values for nvcc and for other versions relative to nvcc) on k40 GPU}
    \label{tab:res_calc_2l_instruction}
\end{table}
\begin{table}
    \centering
    \begin{tabular}{l|c|c|c|P{2cm}|P{2cm}}
         &  fp (64 bit) &integer   &control & Texture read transaction & Global read transaction  \\ \hline
         nvcc - \cuda		 &93555K&94994K&1439K&2175K&334K\\ \hline
         clang - \cuda		 &0.98 &0.94&1.00&1.01&1.04 \\ \hline
         PGI - \acc			 &1.03 &1.38&1.00&0.98&1.00 \\ \hline
         XL - \omp4			 &1.00 &1.50&1.00&0.28&3.53 \\ \hline
         clang - \omp4		 &0.97 &1.26&1.00&0.00&3.42 \\ \hline\hline
         PGI - fortran \cuda &1.03 &1.80&2.00&1.05&13.47\\ \hline
         PGI - fortran \acc	 &1.03 &1.55&1.00&3.73&3.73 \\ \hline
         XL - fortran \cuda	 &1.00 &2.20&2.00&3.74&3.73 \\ \hline
         XL - fortran \omp4  &1.00 &1.82&1.00&3.77&3.73 \\ \hline
    \end{tabular}
    \caption{Average number of instructions and transactions performed in \res\ kernel with global coloring (absolute values for nvcc and for other versions relative to nvcc) on K40 GPU}
    \label{tab:res_calc_glob_instruction}
\end{table}

\begin{table}
	\centering
    \begin{tabular}{l|P{1.25cm}|P{1.05cm}|P{1.9cm}|P{1.25cm}|P{1.05cm}|P{1.9cm}}
    		& \multicolumn{3}{c}{K40} 		& \multicolumn{3}{c}{P100}\\
         	& Run Time (s) & BW (GB/s) & Reg. count (Occupancy) & Run Time (s) & BW (GB/s) & Reg. count (Occupancy)\\ \hline
		nvcc - \cuda\	& 0.064 &32&44 (62.5\%) & 0.032&64&48 (62.5\%)\\ \hline
		clang - \cuda\ 	& 0.064 &32&44 (62.5\%) & 0.032&64&46 (62.5\%)\\ \hline\hline
		PGI - F\_\cuda\ & 0.082 &25&53 (56.25\%)& 0.029&71&72 (43.75\%)\\ \hline
		XL - F\_\cuda\ 	& 0.061 &33&48 (62.5\%) & 0.035&59&64 (50\%) \\ \hline
	\end{tabular}
	\caption{Measured run time, bandwidth, register count and occupancy values in case of \bres\ in case of hierarchical coloring} \vspace{-10pt}
	\label{tab:bres_calc_2l}
    \begin{tabular}{l|P{1.25cm}|P{1.05cm}|P{1.9cm}|P{1.25cm}|P{1.05cm}|P{1.9cm}}
    		& \multicolumn{3}{c}{K40} 		& \multicolumn{3}{c}{P100}\\
         	& Run Time (s) & BW (GB/s) & Reg. count (Occupancy) & Run Time (s) & BW (GB/s) & Reg. count (Occupancy)\\ \hline
		nvcc - \cuda	& 0.072 &28&44 (62.5\%) & 0.035&58&42 (62.5\%)\\ \hline
		clang - \cuda	& 0.071 &29&38 (75\%)   & 0.034&59&37 (75\%)\\ \hline
		PGI - \acc		& 0.072 &28&71 (43.75\%)& 0.034&60&56 (56.25\%)\\ \hline
		XL - \omp4		& 0.084 &24&72 (43.75\%)& 0.037&55&80 (37.5\%)\\ \hline
		clang - \omp4	& 0.079 &26&88 (31.25\%)& 0.039&52&94 (31.25\%)\\ \hline\hline
		PGI - F\_\cuda 	& 0.096 &21&56 (56.25\%)& 0.038&54&72 (43.75\%) \\ \hline
		PGI - F\_\acc	& 0.073 &28&102 (25\%)  & 0.036&57&88 (31.25\%)\\ \hline
		XL - F\_\cuda	& 0.078 &26&70 (43.75\%)& 0.037&55&80 (37.5\%)\\ \hline
		XL - F\_\omp4	& 0.078 &26&94 (31.25\%)& 0.035&57&80 (37.5\%)\\ \hline
	\end{tabular}
	\caption{Measured run time, bandwidth, register count and occupancy values in case of \bres\ in case of global coloring} \vspace{-30pt}
	\label{tab:bres_calc}
\end{table} 

\subsubsection{\bres}:
The \bres\ kernel also has indirect reads and writes, so we need coloring like with \res. In \bres\ the versions using hierarchical coloring performs equally good except the Fortran \cuda\ version compiled with the PGI compiler as shown in Table \ref{tab:bres_calc_2l}. The \cuda\ Fortran version with the PGI compiler has 30\% lower performance compared to other versions with hierarchical coloring. On the K40 GPU in \res\ \cuda\ PGI has high number of load transactions but in this case the PGI version doesn't use the texture cache. However on the P100 GPU the version using the PGI compiler have same amount of memory transactions as nvcc, but executes less floating point operations. 
In case of global coloring on the C/C++ side \acc\ performs as good as \cuda\ versions despite the lower occupancy as shown in Table \ref{tab:bres_calc}. However the \omp4 versions have the same issue as in case of \res\ and get high number of global read transactions while don't use the texture cache, which (with the lower occupancy due to high register counts) leads to the 20\% lower performance. \\
In this case Fortran versions have only 10\% lower performance than C/C++ versions (except for \cuda\ with the PGI compiler which has the same issue as with hierarchical coloring). The key reason for the difference is the lower occupancy of the Fortran versions and the higher instruction and memory transaction counts on both GPU. However in this case the directive based approaches performing equally to \cuda\ Fortran with the XL compiler. Surprisingly for \bres\ the Fortran \acc\ version has as low register count as the \cuda\ versions on the C/C++ side.

\vspace{-10pt}
\subsubsection{\update}:

The \cuda\ Fortran versions have lower occupancy because of the high register usage (the \acc\ version has a separate kernel for reduction thus have lower register count for the bulk of the kernel and the \omp4 version performs about the same as the C/C++ versions). All of the Fortran versions ended up with about 4 times more texture read (except \omp4 which doesn't use texture cache, but has 12 times more global loads), global load and store transactions than \cuda\ with nvcc. \cuda\ Fortran versions also have spilled registers (which introduce about 10k-20k local load and store transactions). 

\begin{table}
    \centering
    \begin{tabular}{l|P{1.25cm}|P{1.05cm}|P{1.9cm}|P{1.25cm}|P{1.05cm}|P{1.9cm}}
    		& \multicolumn{3}{c}{K40} 		& \multicolumn{3}{c}{P100}\\
         	& Run Time (s) & BW (GB/s) & Reg. count (Occupancy) & Run Time (s) & BW (GB/s) & Reg. count (Occupancy)\\ \hline
         nvcc - \cuda		& 4.478&175&31 (100\%)  & 1.519&516&32 (100\%)\\ \hline
         clang - \cuda		& 4.481&175&32 (100\%)  & 1.519&516&32 (100\%)\\ \hline
         PGI - \acc			& 4.416&177&\makecell{36 (75\%) \\ 18 (100\%)}&1.588&493&\makecell{38 (75\%)\\ 12 (100\%)}\\ \hline
         XL - \omp4			& 4.497&174&32 (100\%)  & 1.660&472&32 (100\%)\\ \hline
         clang - \omp4		& 5.175&151&86 (31.25\%)& 1.719&456&86 (31.25\%)\\ \hline\hline
         PGI - F\_\cuda		& 4.598&170&79 (43.75\%)& 1.654&474&48 (62.5\%)	\\ \hline
         PGI - F\_\acc		& 4.350&180&\makecell{37 (75\%)\\ 18 (100\%)}&1.583&495&\makecell{40 (75\%)\\ 16 (100\%)}\\ \hline
         XL - F\_\cuda		& 4.598&169&80 (37.5\%)& 1.566&500&80 (37.5\%)\\ \hline
         XL - F\_\omp4		& 5.074&154&46 (62.5\%)& 1.712&458&40 (75\%)\\ \hline
    \end{tabular}
    \caption{Measured run time, bandwidth, register count and occupancy values in case of \update\ (for \acc\ versions the second register count belongs to the reduction kernel, the run times are the sum of the two kernels)} \vspace{-20pt}
    \label{tab:update}
\end{table}

\vspace{-10pt}
\subsubsection{Effect of tuning the number of registers per thread}

\begin{figure}
\centering \vspace{-20pt}
\includegraphics[width=\linewidth]{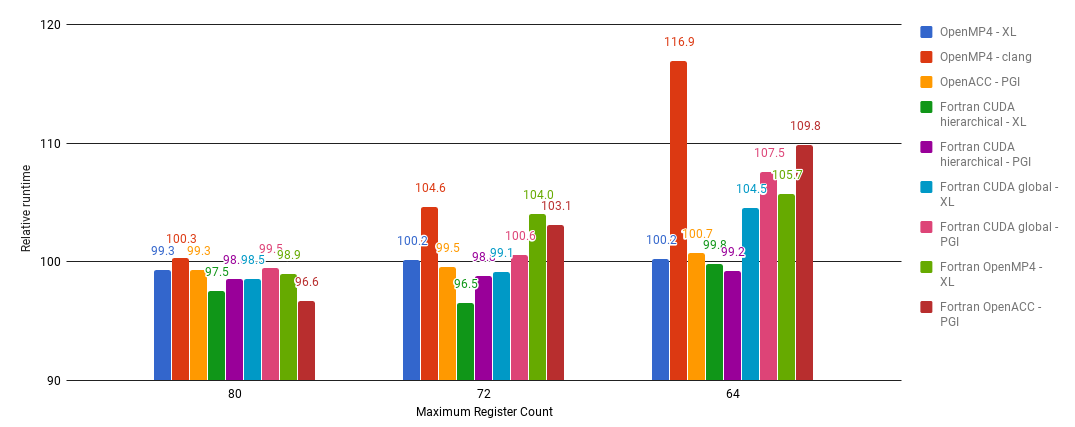}
\caption{Runtime of C OpenACC/OpenMP4 and Fortran versions with limited register per thread relative to original versions measured on K40. Lower is better.}\vspace{-10pt}
\label{fig:regcount_rel_80}
\end{figure}

\begin{figure}
\centering
\includegraphics[width=0.9\linewidth]{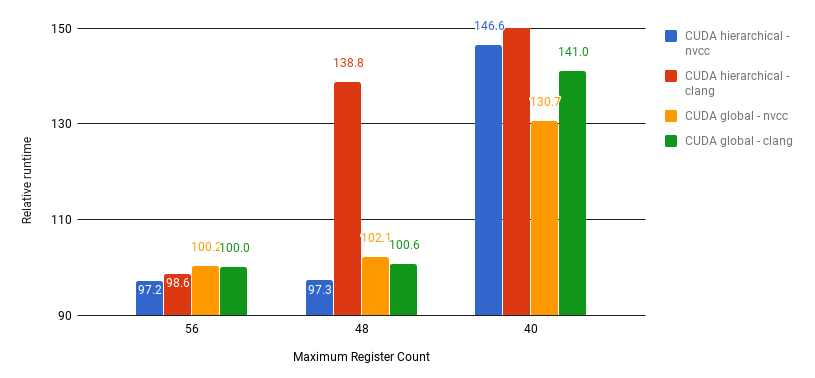}
\caption{Runtime of CUDA C versions with limited register per thread relative to original versions measured on K40. Lower is better.}\vspace{-10pt}
\label{fig:regcount_rel_56}
\end{figure}

In case of the Airfoil application, the key performance limiter is the latency of accesses to global memory. To achieve high bandwidth, we need many loads in flight. This requires increasing the occupancy, which is limited by the number of registers used in these kernels. To get better occupancy we can limit the maximum number of registers per thread during the compilation. The register counts where the occupancy decreases if we use one more register per thread are the same for both K40 and P100 GPUs with 128 thread per block. For \cuda\ C/C++ versions we restricted the register counts to 56, 48 and 40 in order to increase occupancy, while for other versions we got higher register counts thus the restricted the register usage to 80, 72 and 64. With hierarchical coloring the shared memory required by the kernel could be the bottleneck for occupancy. In Figure \ref{fig:regcount_rel_80} and \ref{fig:regcount_rel_56} the runtime of limited versions relative to the original version in percentage are shown. The shared memory requirement of \res\ and \bres\ is roughly 4KB per block which limits the occupancy at 68.8\% on the K40, meaning that we cannot reach better occupancy by further reducing the maximum register count (reducing the count to 48 would lead to 62.5\% and to 40 would lead to 75\% occupancy). On the P100 GPU shared memory requirement maximizes the occupancy at 94\% thanks to more available shared memory. For most language-compiler combinations, limiting the register count only affects the \adt, \res\ and \bres\ kernels. In the \omp4 - clang, Fortran \omp4 - XL, and Fortran \cuda\ - PGI combinations, \update\ is also affected by the limiting because of the high register count as shown in Table \ref{tab:update}. 


With the increased occupancy, we do get better run times in most cases (a limit of 56 in case of C/C++ and \cuda\ and 80 for other versions), except for the clang \omp4 and \cuda\ with nvcc. However further limitation of register counts leads to performance degradation, with the exception of CUDA Fortran code compiled with XL (which have the best performance with register count limited to 72). The reason for the loss of performance is the  increasing number of spilled registers, and the latency introduced by the usage of these registers. 

The main differences lie in the run times of \res\ and \adt. For \res\ on C/C++ side limiting the register count increases the performance by 2-5\% in case of \cuda\ with hierarchical coloring, the \omp4 XL compiler and the Fortran versions also get better run times by 1-2\% but the \acc version performs the same, while the \omp4 clang versions get 2\% higher run time thus get higher total run time despite of the 5\% performance increase in \update\ and \adt. For Fortran \cuda\ with XL and Fortran \acc\ with PGI compiler reach 15\% better performance in \adt\ for the first level limitation. These results implies that for the most cases the increased occupancy gained with the restriction of the register usage could increase performance significantly (especially for kernels with low occupancy). In terms of instruction counts the limitation of register usage leads to slightly increased integer instruction counts in our cases.

\vspace{-10pt}
\subsection{Volna}\vspace{-5pt}

\begin{figure}[t]
	\centering \vspace{-20pt}
    	\includegraphics[width=0.8\textwidth]{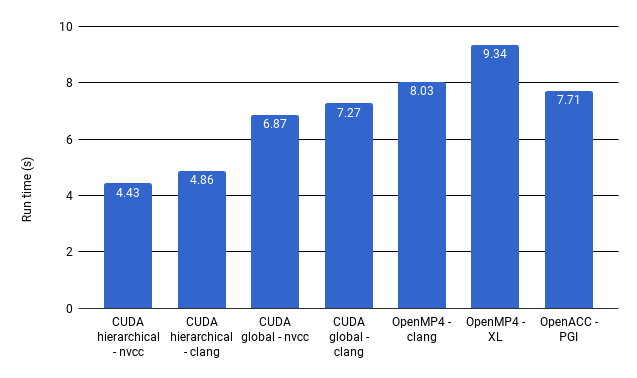}
    	\includegraphics[width=0.8\textwidth]{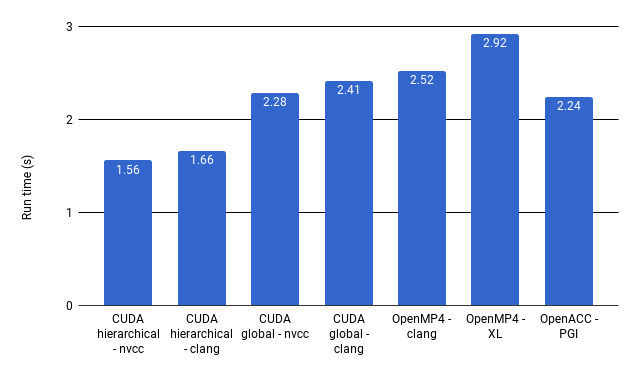}
	    \caption{Measured run times of Volna versions on the K40 and P100 GPU}\label{fig:P100_volna_RT}\vspace{-20pt}
	    
\end{figure}
\begin{table}
\vspace{-10pt}
\begin{tabular}{P{.22\linewidth}|M{.095\linewidth}|M{.1\linewidth}|M{.08\linewidth}|M{.08\linewidth}|M{.12\linewidth}|M{.12\linewidth}|M{.12\linewidth}}
&nvcc \cuda\ &clang \cuda\ &nvcc \cuda\ global&clang \cuda\ global&clang \omp4 &XL \omp4&\makecell{PGI\\ \acc} \\ \hline
\textbf{compute Fluxes}		 	&1.336&1.693&1.323&1.734&2.186&1.613&1.623\\ \hline
\textbf{Space Discretization} 	&1.758&1.834&4.150&4.134&3.973&6.261&4.762\\ \hline
\textbf{Numerical Fluxes}	 	&0.431&0.431&0.507&0.511&0.549&0.528&0.496\\ \hline
\textbf{Evolve Values RK2\_2}	&0.312&0.313&0.326&0.325&0.416&0.300&0.302\\ \hline
\textbf{Evolve Values RK2\_1}	&0.371&0.372&0.366&0.365&0.648&0.383&0.338\\ \hline
\end{tabular}
\caption{Run times of the five most time consuming Volna kernels on the K40 GPU}\vspace{-20pt}
\label{tab:volna}
\end{table}

\begin{table}
\begin{tabular}{P{.21\linewidth}|M{.1\linewidth}|M{.1\linewidth}|M{.085\linewidth}|M{.085\linewidth}|M{.12\linewidth}|M{.12\linewidth}|M{.12\linewidth}}
&nvcc \cuda\ &clang \cuda\ &nvcc \cuda\ global&clang \cuda\ global&clang \omp4 &XL \omp4&\makecell{PGI\\ \acc} \\ \hline
\textbf{compute Fluxes}		 	&56 (56.25\%)&60 (56.25\%)&22 (100\%) &22 (100\%) &93 (31.25\%) &78 (37.5\%)&77 (37.5\%)\\ \hline
\textbf{Space Discretization} 	&32 (100\%)	 &36 (75\%)	  &28 (100\%) &25 (100\%) &64 (50\%)	&30 (100\%) &30 (100\%)\\ \hline
\textbf{Numerical Fluxes}	 	&28 (100\%)	 &16 (100\%)  &45 (62.5\%)&46 (62.5\%)&40 (75\%)	&30 (100\%) &\makecell{33 (75\%)\\12  (100\%)}\\ \hline
\textbf{Evolve Values RK2\_2}	&26 (100\%)	 &24 (100\%)  &26 (100\%) &24 (100\%) &80 (37.5\%)	&25 (100\%) &28 (100\%)\\ \hline
\textbf{Evolve Values RK2\_1}	&28 (100\%)	 &27 (100\%)  &28 (100\%) &27 (100\%) &86 (31.25\%)	&32 (100\%) &33 (75\%)\\ \hline

\end{tabular}
\caption{Register counts and occupancy of the five most time consuming Volna kernels on the K40 GPU (for \acc\ the second register count belongs to the reduction kernel)}
\label{tab:volna_RC}
\end{table}
For Volna the \textbf{SpaceDiscretization} kernel has a huge impact on runtime (half of the time is spent in this kernel when using global coloring), and so the hierarchical coloring leads to significant overall performance gain as shown on Figure \ref{fig:P100_volna_RT} (the measurements are in single precision because Volna requires only single precision to get correct results). However the presence of the local reads in \textbf{computeFluxes} in case of clang \cuda\ leads to 20\% performance loss in this kernel. On other kernels we found the same tendencies as we observed on Airfoil, i.e clang reaches lower floating point and integer instruction counts compared to nvcc. The directive based approaches have lower performance in the two most time consuming kernels. The \omp4 with XL has about 50\% lower performance in \textbf{SpaceDiscretization} on the K40 GPU (the difference is 40\% on the P100 machine), while for the other kernels these approaches performed within 10\% of \cuda's performance and in some cases even better as shown in Table \ref{tab:volna}. In terms of occupancy the \omp4 with XL reach about the same occupancy in most cases as \cuda, while \omp4 clang and \acc\ have high register counts as shown in Table \ref{tab:volna_RC}. In terms of instruction counts in case of Volna the directive based approaches performed the same as in Airfoil. The \omp4 versions don't use texture caches (in case of XL the texture cache usage is about 15\% of nvcc's) and all directive based approach have higher global read transactions and about 30\% higher integer instruction counts.

\subsection{BookLeaf}
\begin{figure}
	\centering
    	\includegraphics[width=0.8\textwidth]{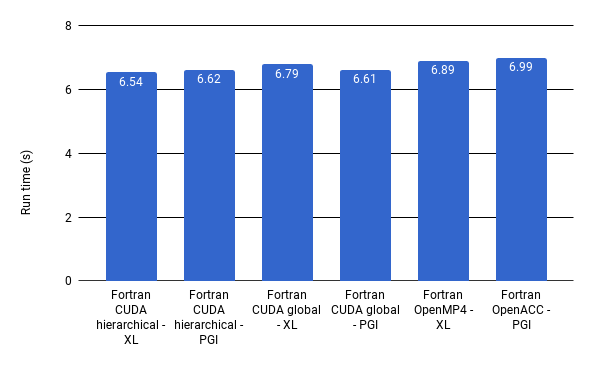}
    	\includegraphics[width=0.8\textwidth]{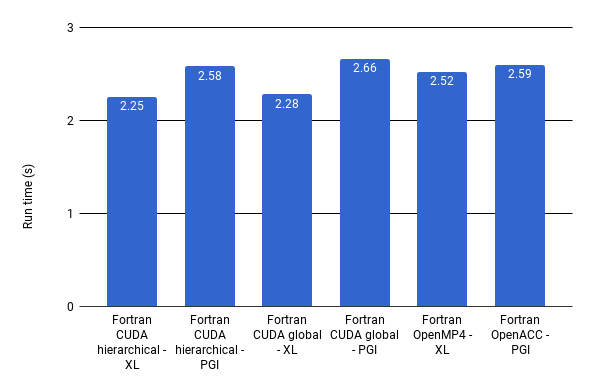}
	    \caption{Measured run times of BookLeaf versions on K40 and P100 GPU}
	    \label{fig:k40_bookleaf_RT} 
\end{figure}

Considering that in BookLeaf most of the time is spent in direct kernels or indirect read kernels, there is not as much difference between hierarchical and global coloring versions in total run time, as shown in Figure \ref{fig:k40_bookleaf_RT}. However in case of \textbf{getacc\_scatter}, which is the only kernel with indirect increments among the top five most time consuming kernels, the runtime of the hierarchical coloring is at least 50\% better than that of the global coloring versions. All versions are within 7\% of the performance of the best version which is Fortran \cuda\ with hierarchical coloring compiled with the XL compiler on the K40, while on the P100 machine the PGI compiler performance is about 10\% lower than the performance of the versions compiled with the XL compiler. As we saw in Airfoil, the \cuda\ versions have high register count in the most cases, but \acc\ and \omp4 reach better occupancy as shown in Table~\ref{tab:bookleaf_RC} which leads even better runtime than in case of \cuda\ versions as shown in Table~\ref{tab:bookleaf}.\\

\vspace{-20pt}
\begin{table}\vspace{-10pt}
\begin{tabular}{l|M{.11\linewidth}|M{.11\linewidth}|M{.12\linewidth}|M{.11\linewidth}|M{.11\linewidth}|M{.12\linewidth}}
&\cuda\ - PGI&\cuda\ global - PGI&\acc&\cuda\ - XL&\cuda\ global - XL&OpenMP4\\ \hline
\textbf{getq\_christiensen1}	&0.937&0.937&1.033&0.979&0.987&0.866\\ \hline
\textbf{getq\_christiensen\_q}	&0.933&0.934&0.975&0.888&0.889&0.751\\ \hline
\textbf{getacc\_scatter}		&0.457&0.450&0.917&0.497&0.785&0.769\\ \hline
\textbf{gather}					&0.526&0.526&0.525&0.523&0.523&0.542\\ \hline
\textbf{getforce\_visc}			&0.493&0.493&0.484&0.421&0.421&0.390\\ \hline
\end{tabular}
\caption{Run times of the five most time consuming BookLeaf kernel on K40 GPU}\vspace{-10pt}
\label{tab:bookleaf}
\begin{tabular}{l|M{.11\linewidth}|M{.11\linewidth}|M{.12\linewidth}|M{.11\linewidth}|M{.11\linewidth}|M{.12\linewidth}}
&\cuda\ - PGI&\cuda\ global - PGI&\acc&\cuda\ - XL&\cuda\ global - XL&OpenMP4\\ \hline
\textbf{getq\_christiensen1}	&78 (37.5\%) &78 (37.5\%) &77 (37.5\%)  &144 (18.75\%)&86 (31.25\%)&78 (37.5\%)\\ \hline
\textbf{getq\_christiensen\_q}	&86 (31.25\%)&86 (31.25\%)&143 (18.75\%)&126 (25\%)	  &126 (25\%)  &70 (43.75\%)\\ \hline
\textbf{getacc\_scatter}		&75 (37.5\%) &79 (37.5\%) &28 (100\%)   &96 (31.25\%) &54 (56.25\%)&23 (100\%)\\ \hline
\textbf{gather}					&30 (100\%)  &30 (100\%)  &23 (100\%) 	&32 (100\%)	  &32 (100\%)  &23 (100\%)\\ \hline
\textbf{getforce\_visc}			&44 (62.5\%) &40 (75\%)   &32 (100\%)	&56 (56.25\%) &56 (56.25\%)&32 (100\%)\\ \hline
\end{tabular}
\caption{Register counts of the five most time consuming BookLeaf kernel on K40 GPU}\vspace{-30pt}
\label{tab:bookleaf_RC}
\end{table}


\section{Conclusions} \label{sec/conc}
In this paper we have carried out a detailed study of some of the most popular parallelization approaches, programming languages, and compilers used to program GPUs, on a number of parallel loops coming from the domain of unstructured mesh computations. \omp4 and \acc\ are high level models using directives on loops in order to utilize GPUs, while \cuda\ use a lower level Single Instruction Multiple Threads model.

In this class of applications, a key common computational pattern is the indirect incrementing of data: to avoid race conditions we explored the use of coloring. The high level models must use global coloring of the iteration set to ensure that no two threads writes the same value when running simultaneously, whereas with lower-level models (CUDA) it is possible to apply a ``two-level'' coloring approach permitting better data reuse.

In case of Fortran, the \cuda\ versions with global coloring and \acc\ versions are within 10\% of each other's performance. However the \omp4 versions use higher number of registers per thread in some cases, leading to low occupancy, as well as lower performance executing reductions. Directive based approaches also use higher numbers of integer and control instructions.

On the C/C++ side, \cuda\ code compiled with the clang compiler performs 2-5\% better in terms of runtime and in most cases can outperform nvcc in the optimization of computations thus perform 20\% fewer integer and floating point instructions compared to nvcc. The higher level approaches currently using more registers (even for simple kernels in case of \omp4 with the clang compiler) which leads to lower occupancy that lowers the performance. Also these versions now executing 30\% more integer instructions than \cuda, but in some cases they performs within 5\% of nvcc's  performance. Since the support for \omp4 is relatively new there are still some issues that lowers performance, such as the more infrequent use of the texture cache and the lower performance when performing reductions. Also the \acc\ and \omp4 with the XL compiler currently have problems with computations with multiple increment of the same data as in \adt\ where these versions write back all intermediate result to the global memory introducing the gap between the their performance and \cuda's.

We have also shown that using \cuda\ one can handle race conditions more efficiently thanks to block-level synchronization; this in turn enables an execution approach with much higher data reuse. Kernels with indirect increments using hierarchical coloring have significantly better performance than the versions using global coloring; in case of Airfoil hierarchical coloring leads to about 35\% better overall performance, for Volna the difference is about 50\% and with BookLeaf about 3\%.

In summary, we have demonstrated that support for C is only slightly better than for Fortran, for all possible combinations, with a 3-10\% performance gap. Our work is among the first ones comparatively evaluating the clang CUDA compiler and IBM's XL compilers; clang's CUDA support is showing great performance already, often outperforming nvcc. Even though the XL compilers are only about one year old, they are already showing competitive performance and good stability - on the OpenMP 4 side often outperforming clang's OpenMP 4 and PGI's OpenACC. Directive based approaches demonstrate good performance on simple computational loops, but struggle with more complex kernels due to increased register pressure and instruction counts - lagging behind CUDA on average by 5-15\%, but in the worst cases by up to 50\%. It still shows that OpenMP 4 GPU support isn't yet as mature as OpenACC, nevertheless, they are within 5-10\%. Our results also demonstrate how CUDA allows for more flexibility in applying optimizations that are currently not possible with OpenACC or OpenMP 4.

\section*{Acknowledgements}
The authors would like to thank the IBM Toronto compiler team, and Rafik Zurob in particular, for access to beta compilers and help with performance tuning, and Michal Iwanski and J\'ozsef Sur\'anyi at IBM for access to a Minsky system. Thanks to Carlo Bertolli at IBM TJ Watson for help with the clang OpenMP 4 compiler. This paper was supported by the J\'anos B\'olyai Research Scholarship of the Hungarian Academy of Sciences. The authors would like to acknowledge the use of the University of Oxford Advanced Research Computing (ARC) facility in carrying out this work \url{http://dx.doi.org/10.5281/zenodo.22558}. The research has been supported by the European Union, co-financed by the European Social Fund (EFOP-3.6.2-16-2017-00013).

%
%
\bibliographystyle{splncs}
\bibliography{main}

\begin{thebibliography}{10}

\bibitem{cuda}
Nickolls, J., Buck, I., Garland, M., Skadron, K.:
\newblock Scalable parallel programming with cuda.
\newblock Queue \textbf{6}(2) (March 2008)  40--53

\bibitem{opencl}
Stone, J.E., Gohara, D., Shi, G.:
\newblock Opencl: A parallel programming standard for heterogeneous computing
  systems.
\newblock IEEE Des. Test \textbf{12}(3) (May 2010)  66--73

\bibitem{openacc}
Wienke, S., Springer, P., Terboven, C., an~Mey, D.:
\newblock Openacc: First experiences with real-world applications.
\newblock In: Proceedings of the 18th International Conference on Parallel
  Processing. Euro-Par'12, Berlin, Heidelberg, Springer-Verlag (2012)  859--870

\bibitem{openmp4}
:
\newblock {OpenMP 4.5 specification}.
\newblock \url{http://www.openmp.org/wp-content/uploads/openmp-4.5.pdf}

\bibitem{Wu:2016:GOG:2854038.2854041}
Wu, J., Belevich, A., Bendersky, E., Heffernan, M., Leary, C., Pienaar, J.,
  Roune, B., Springer, R., Weng, X., Hundt, R.:
\newblock {Gpucc: An Open-source GPGPU Compiler}.
\newblock In: Proceedings of the 2016 International Symposium on Code
  Generation and Optimization. CGO '16, New York, NY, USA, ACM (2016)  105--116

\bibitem{pgi}
:
\newblock {The Portland Group}.
\newblock \url{http://www.pgroup.com}

\bibitem{ruetsch2013cuda}
Ruetsch, G., Fatica, M.:
\newblock CUDA Fortran for scientists and engineers: best practices for
  efficient CUDA Fortran programming.
\newblock Elsevier (2013)

\bibitem{xlcuf}
:
\newblock {Getting Started with CUDA Fortran programming using XL Fortran for
  Little Endian Distributions}.
\newblock
  \url{http://www-01.ibm.com/support/docview.wss?uid=swg27047958&aid=11}

\bibitem{clang-ykt}
:
\newblock {Clang with OpenMP 4 support}.
\newblock \url{https://github.com/clang-ykt}

\bibitem{ledur2013comparative}
Ledur, C.L., Zeve, C.M., dos Anjos, J.C.:
\newblock Comparative analysis of openacc, openmp and cuda using sequential and
  parallel algorithms.
\newblock In: 11th Workshop on parallel and distributed processing (WSPPD).
  (2013)

\bibitem{herdman2012accelerating}
Herdman, J., Gaudin, W., McIntosh-Smith, S., Boulton, M., Beckingsale, D.A.,
  Mallinson, A., Jarvis, S.A.:
\newblock Accelerating hydrocodes with openacc, opencl and cuda.
\newblock In: High Performance Computing, Networking, Storage and Analysis
  (SCC), 2012 SC Companion:, IEEE (2012)  465--471

\bibitem{6546071}
Hoshino, T., Maruyama, N., Matsuoka, S., Takaki, R.:
\newblock Cuda vs openacc: Performance case studies with kernel benchmarks and
  a memory-bound cfd application.
\newblock In: 2013 13th IEEE/ACM International Symposium on Cluster, Cloud, and
  Grid Computing. (May 2013)  136--143

\bibitem{norman2015case}
Norman, M., Larkin, J., Vose, A., Evans, K.:
\newblock A case study of cuda fortran and openacc for an atmospheric climate
  kernel.
\newblock Journal of computational science \textbf{9} (2015)  1--6

\bibitem{kuan2014accelerating}
Kuan, L., Neves, J., Pratas, F., Tom{\'a}s, P., Sousa, L.:
\newblock Accelerating phylogenetic inference on gpus: an openacc and cuda
  comparison.
\newblock In: IWBBIO. (2014)  589--600

\bibitem{gong2016nekbone}
Gong, J., Markidis, S., Laure, E., Otten, M., Fischer, P., Min, M.:
\newblock Nekbone performance on gpus with openacc and cuda fortran
  implementations.
\newblock The Journal of Supercomputing \textbf{72}(11) (2016)  4160--4180

\bibitem{antao2016offloading}
Antao, S.F., Bataev, A., Jacob, A.C., Bercea, G.T., Eichenberger, A.E., Rokos,
  G., Martineau, M., Jin, T., Ozen, G., Sura, Z.,  et~al.:
\newblock Offloading support for openmp in clang and llvm.
\newblock In: Proceedings of the Third Workshop on LLVM Compiler Infrastructure
  in HPC, IEEE Press (2016)  1--11

\bibitem{Martineau2016}
Martineau, M., Price, J., McIntosh-Smith, S., Gaudin, W.
\newblock In: Pragmatic Performance Portability with OpenMP 4.x. Springer
  International Publishing, Cham (2016)  253--267

\bibitem{Karlin2016}
Karlin, I., Scogland, T., Jacob, A.C., Antao, S.F., Bercea, G.T., Bertolli, C.,
  de~Supinski, B.R., Draeger, E.W., Eichenberger, A.E., Glosli, J., Jones, H.,
  Kunen, A., Poliakoff, D., Richards, D.F.
\newblock In: Early Experiences Porting Three Applications to OpenMP 4.5.
  Springer International Publishing, Cham (2016)  281--292

\bibitem{Hart2015FirstEP}
Hart, A.:
\newblock First experiences porting a parallel application to a hybrid
  supercomputer with openmp4.0 device constructs.
\newblock In: IWOMP. (2015)

\bibitem{reguly2016high}
Reguly, I.Z., Keita, A.K., Zurob, R., Giles, M.B.:
\newblock High performance computing on the ibm power8 platform.
\newblock In: International Conference on High Performance Computing, Springer
  (2016)  235--254

\bibitem{op2}
:
\newblock {OP2 github repository}.
\newblock \url{https://github.com/OP2/OP2-Common}

\bibitem{mudalige2012op2}
Mudalige, G., Giles, M., Reguly, I., Bertolli, C., Kelly, P.:
\newblock Op2: An active library framework for solving unstructured mesh-based
  applications on multi-core and many-core architectures.
\newblock In: Innovative Parallel Computing (InPar), 2012, IEEE (2012)  1--12

\bibitem{giles2011performance}
Giles, M.B., Mudalige, G.R., Sharif, Z., Markall, G., Kelly, P.H.:
\newblock Performance analysis and optimization of the op2 framework on
  many-core architectures.
\newblock The Computer Journal \textbf{55}(2) (2011)  168--180

\bibitem{giles2012op2}
Giles, M., Mudalige, G., Reguly, I.:
\newblock Op2 airfoil example.
\newblock (2012)

\bibitem{dutykh2011volna}
Dutykh, D., Poncet, R., Dias, F.:
\newblock The volna code for the numerical modeling of tsunami waves:
  Generation, propagation and inundation.
\newblock European Journal of Mechanics-B/Fluids \textbf{30}(6) (2011)
  598--615

\bibitem{uk-mac}
:
\newblock {Uk mini-app consortium}.
\newblock \url{https://uk-mac.github.io}

\end{thebibliography}

\end{document}